\documentstyle[12pt,aasms4]{article}

\lefthead{C. Gibson and R. Schild}
\righthead{Quasar-microlensing versus
star-microlensing}

\begin{document}

\title{
   Quasar-microlensing versus star-microlensing
evidence of small-planetary-mass objects as the
dominant inner-halo galactic dark matter}

\author{Carl H. Gibson}
\affil{Mechanical and Aerospace Engineering and
Scripps Institution\\ of Oceanography Departments,
University of California,
   San Diego, CA 92093-0411; \\ cgibson@ucsd.edu}

\and

\author{Rudolph E. Schild}
\affil{Center for Astrophysics,
    60 Garden Street, Cambridge, MA 02138; \\
rschild@cfa.harvard.edu}

\begin{abstract} We examine recent results of two
kinds of microlensing experiments intended to
detect galactic dark matter objects, and we suggest
that the lack of short period  star-microlensing
events observed for stars near the Galaxy does not
preclude either the ``rogue planets''  identified 
from quasar-microlensing by Schild 1996  as the
missing-mass of a lens galaxy, or the ``Primordial
Fog Particles'' (PFPs) in
Proto-Globular-star-Cluster (PGC) clumps predicted
by Gibson
$1996\--2000$ as the dominant inner-halo galactic
dark matter component from a new hydrodynamic
gravitational structure formation theory.  We point
out that hydro-gravitational processes acting on a
massive population of such micro-brown-dwarfs
in  their nonlinear accretional cascades to form
stars gives intermittent lognormal number density
$n_p$ distributions for the PFPs within the
PGC gas-stabilized-clumps.
Hence, star-microlensing  searches that focus on a
small fraction of the sky assuming a uniform
distribution for $n_p$ are subject to vast
underestimates of the mean 
${\langle n_p \rangle}_{mean}$.  Sparse independent
samples give modes  
$10^{-4} \-- 10^{-6}$ smaller than means of the
highly skewed lognormal distributions expected. 
Quasar-microlensing searches with higher optical
depths are less affected by $n_p$ intermittency. We
attempt to reconcile the results of the
star-microlensing and quasar-microlensing studies,
with particular reference to the necessarily
hydrogenous and primordial small-planetary-mass
range. We conclude that star microlensing searches
cannot exclude and are unlikely even to detect these
low-mass candidate-galactic-dark-matter-objects so
easily  observed by quasar-microlensing and so
robustly predicted by the new theory.

\end{abstract}

\keywords{cosmology: theory, observations --- dark
matter --- Galaxy:  halo --- gravitational lensing
--- turbulence}

\section{Introduction}

Clearly the missing-mass problem is central to
astronomy today since all attempts to understand
the growth of structure in the universe, as well as
the dynamics of galaxies and their clusters,
require an understanding of the nature and location
of this dominant mass component (\cite{car94}). 
Because most of the inner-halo galactic missing-mass
is probably in compact objects,  attempts to detect
it have been based upon quasar-microlensing  and
star-microlensing which can detect objects of any
mass but cannot detect a smooth
distribution of matter.

The first published quasar-microlensing reports were
from lensed quasars by 
\cite{van89} and \cite{irw89}, although certainly
the high optical depth Q0957+561A,B
quasar-microlensing reported by Vanderriest et al.
was known by
\cite{sch86}  who did not presume to discuss
microlensing until their measured time delay
difference of 1.1 years between the A and B quasar
images was confirmed.  A secure image time delay
difference is required to allow correction for
intrinsic fluctuations of the source brightness at
$0.1\--0.3$ year planetary-mass periods by
subtraction of the image light curves matched at
the time their light was lensed.  The 1.1 year
\cite{sch86} delay value has been confirmed and is
now generally accepted,
\cite{kun97}. 
Star-microlensing searches at low optical depth in
directions of large star densities toward the Large
Magellanic Cloud (LMC) and the Galactic Bulge
resulted in detections (\cite{alc93}, 
 \cite{aub93}, \cite{uda93}). However the
interpretation of the microlensing statistics from
the two kinds of programs has produced disagreement.

The quasar-microlensing in Q2237 was first to be
analyzed, and earliest reports suggesting solar
mass stars (\cite{wam90}, 
\cite{cor91}, \cite{yee92}) were later amended by
\cite{ref93} who demonstrated that the apparently
rapid fluctuations detected in the observations
could indicate a dominant population of terrestrial
mass microlenses,  but called for more accurate
lightcurves over longer time spans to make the
indication conclusive.  Interpretation of the Q0957
microlensing was delayed by a long controversy
about the time delay, but the critical fact that
rapid microlensing is observed was already noted by
\cite{sch91}. Exhaustive statistical analyses of
the Q0957 data  caused \cite{sch95}
to again conclude that rapid microlensing is
observed, and a power spectrum by \cite{tho96}, 
also shown in
\cite{sch96}, led to the Schild conclusion that the
microlensing is caused by ``rogue planets ...
likely to be the missing mass.''   A primordial
origin for such micro-brown-dwarf (MBD) objects  as
the baryonic galactic dark matter had already been
proposed by \cite{gib96}, who predicted a 
condensation mass of about 
$10^{-7}$ $M_{\sun}$ for the relatively inviscid
gas formed at photon decoupling from the cooling
viscous plasma.  Recent best estimates from fossil
turbulence theory by
\cite{gib00a}bc of $10^{-6}$ $M_{\sun}$ for the
``Primordial Fog Particle'' (PFP) mass are
closer to the ``rogue planet'' value found by Schild
1996 for Q0957, and are compatible with
results for Q2237 as tentatively interpreted by
\cite{ref93}.  The \cite{gib96} theory
predicts simultaneous fragmentation of
the primordial gas into
$10^{6}$ $M_{\sun}$ proto-globular-cluster (PGC)
clumps as expected from the
\cite{jns02} theory (but for different reasons), 
adding yet another source of unaccounted for
undersampling error to the star-microlensing
searches.

Microlensing of stars has to date produced many
detections, but their interpretation is clouded by
uncertainties in the models of the Halo of the
Galaxy. The original theoretical work that
justified these searches (\cite{pac86},
\cite{gri91})  did not take into account that
microlensing of LMC stars might be produced by dark
matter or foreground stars on the near side of the
LMC itself, as pointed out by \cite{sah94}. This
produces uncertainties in the interpretation of
star-microlensing so large that parallax
observations from one or more satellites may be
needed to accurately compute  the locations and
parameters of the lenses (\cite{gau97}). Moreover,
it was anticipated in the MACHO project
experimental design  (\cite{grt91})
that any Halo  dark matter objects would have mass
$(10^{-3}-10^{-1}) \, M_{\sun}$ much larger than 
$10^{-7} M_{\sun}$, so search frequencies and
strategies prevented detection of lower mass
objects for the first several years. In all
MACHO/EROS calculations of exclusion estimates a
spatially uniform distribution is assumed, rather 
than the highly intermittent distributions expected 
($\S$\ref{sec2}) if the Halo dark matter mass is
dominated by hydrogenous planetoids  (PFPs, MBDs)
in dark PGC clumps.  A small fraction of the LMC
stars sampled may intersect PGCs, but even these
have little chance of  any PFP microlensing within
practical observational periods from the extremely
intermittent PFP number density
$n_p$ we show will occur within most PGCs due to
their accretional cascades.

Whereas the available statistics of
star-microlensing do not appear to have
convincingly produced a dark matter detection
(\cite{ker98}), they have sometimes, but not
consistently, been presented as precluding the
population discovered in quasar-microlensing. The
purpose of the present manuscript is to compare
these two microlensing approaches to the
missing-mass detection problem, with particular
reference to the PGC clumps of PFP planetoids  
proposed by \cite{gib96} as the dominant inner-halo
component of the galaxy missing-mass based on
nonlinear (non-Jeans) fluid mechanical theory (\S
\ref{sec1}).  We show in our Conclusions (\S
\ref{secc})  that results from quasar-microlensing
observations  (\S \ref{seca}) and star-microlensing
observations (\S \ref{secb})
 are actually compatible, considering the
intermittent lognormal number density expected for
these objects as they cluster in a nonlinear
cascade to form larger clusters,  larger objects
and finally, for a small percentage, stars (\S
\ref{sec2}).  

\section{Quasar-microlensing observations}
\label{seca}

The first dark matter conclusion based upon
microlensing was by \cite{kee82}, who confirmed
from 80 years of archived images  the remarkable
concentration of bright quasars near  bright
galaxies noted by \cite{bur79} and \cite{arp80}.
Keel found QSO surface densities $\sigma$ increased
by more than two orders of magnitude at separations
 $R \approx$  10 kpc on the galaxy plane above
background values  with separations $R \ge$ 100
kpc,   but showed low amplitudes of quasar
brightness fluctuations  (less than 1 magnitude or
17\%) at stellar-mass frequencies.  Keel therefore
concluded that ``if gravitational amplification is
involved in producing an excess of bright QSOs near
galaxies, ordinary halo stars are not
responsible.'' The effective radius of the bright
QSO density distribution was 17 kpc, ``slightly
larger than globular cluster systems of the Galaxy
and M31.'' From this evidence, the mass of the
microlensing objects must either be much larger
than $M_{\sun}$ or much smaller, and concentrated
within the $10^{21}$ m (32 kpc) inner-galaxy-halo
range observed for globular clusters around
galaxies.

The directly measured microlensing  (by
\cite{van89})  for Q0957 for a 415 day time delay
was not analyzed for missing-mass implications for
many years, possibly in part because the time delay
would remain controversial.  The detected
microlensing in Q2237, first reported in
\cite{irw89} and initially interpreted as due to
masses in the range 
$0.001 M_{\sun} < M < 0.1 M_{\sun}$, was
reinterpreted in terms of a statistical
microlensing  theory by \cite{ref93}, applicable to
the case of  resolved accretion disks, giving
possible microlensing masses as small as $10^{-7}
M_{\sun}$.

With time delay issues for the Q0957 system
resolved in favor of the original (\cite{sch86})
time delay value, the microlensing was
 discussed by \cite{sch94}, 1995, who removed the
intrinsic source fluctuations of the A image by
subtracting the B image shifted 404 days forward, 
revealing a continuously fluctuating pattern in the 
difference signal, due only to microlensing, that
would be evidence of the missing-mass. A Fourier
power spectrum of these microlensing fluctuations
in \cite{sch95}, and updated in \cite{sch96}, shows
that the microlensing power spectral area is
dominated by a broad peak whose center frequency
corresponds to a microlensing mass of $10^{-5.5}
M_{\sun}$ ($4.4\times10^{24}$ kg).  The dominant
microlens mass might be smaller than this due to
clumping, but not much larger.

To understand why quasar-microlensing and
star-microlensing experiments seem to give
different results, it is necessary to understand
how the higher optical depth in quasar-microlensing
affects the outcome  differently for
 microlensing masses with the intermittent spatial
distribution expected for any primordial population
of small self-gravitating masses.  In
quasar-microlensing, two or more images of the same
quasar are observed, and a normal galaxy, acting as
a lens, will produce unit optical depth to
microlensing. The definition of optical depth is
given precisely in \cite{schn92}.  It is formulated
for an unresolved source, and is  effectively the
probability that a microlensing event is underway.
The surface optical depths for the (A,B) images of
the Q0957 system are (0.3, 1.3), which means that
for image B it is virtually certain that at least
one microlensing event is underway at any time. 
For image A, the probability is only about 30\%
that at any statistical moment a microlensing event
is underway. These probabilities are independent of
the microlensing mass, which can be estimated from
the event duration.

A more complex situation, more favorable for the
detection of microlensing, exists for the Q0957
system, because the Einstein rings of all
microlensing masses are substantially smaller than
the size of the luminous quasar accretion disk.
\cite{sch96} used the historic record of Q0957
brightness fluctuations, in combination with the
appropriate \cite{ref93} statistical theory, to
estimate the size of the quasar source as 6 times
larger than the stellar-microlens Einstein ring. 
For a half-solar mass typical microlens, presumed
to be present from the lens galaxy's spectrum, this
gives an accretion disk diameter of $3 \times
10^{15}$ m.  More importantly, this result means
that at any time there are $6^2$ independent lines
of sight to the quasar B image that each has a
probability of   1.3 of hosting a statistically
independent stellar-microlensing test, or a net
probability of $1.3 \times 6^2$ stellar mass 
microlenses at any moment, and the observed
brightness fluctuations  on a 30-year time scale
should simply reflect the Poisson statistics of
this number of occupied stellar microlensing sites.

The luminous matter described above is known to
exist from the lens galaxy's light, whose spectrum
is dominated by stars of approximately half-solar
mass. From the amplitudes and durations of its
stellar-microlensing  events we have made
inferences about the area of the luminous quasar
source. But \cite{sch96} emphasizes  the
microlensing signature of a dominant second
population of objects, whose 10 to 100 day
fluctuation pattern corresponds to a microlens mass
of
$10^{-6}$ $M_{\sun}$. For these objects to be the
missing-mass, there would need to be at least
$10^{6}$ of them for each half-solar mass star with
the usual assumption that the lens galaxy is
dominated by the missing-mass at the 90\% level or
more. Consistent with the estimates for the quasar
source luminous area given above, the line of sight
to image B should have
$1.3 \times 36 \times 10^{7}$ micro-brown-dwarfs,
and image A should have $0.3 \times 36 \times
10^{7}$, out of $\approx 10^{17}$ in a galaxy. The
brightness fluctuations observed should reflect the
statistical distribution in the numbers of these
microlenses on time scales appropriate for such
small particles (10 to 100 days). The pattern of
brightness fluctuations  in each quasar image
should be continuous and should be symmetrical
(there should be equal positive and negative
fluctuations), as observed in both the A and B
images, and in their phased difference. We discuss
in \S \ref{sec2} the implications of this large
number of statistical tests on estimates of the
statistical parameters of a primordial population
of missing-mass objects. 

A reanalysis of the time delay and microlensing in
this lens system by
\cite{pel98} leads to a smaller source size, but
this does not affect the conclusion that many MBDs
would be projected against the luminous quasar
source at all times. It appears likely that the
microlensing interpretation is beginning to
confront the real quasar source structure, and that
the simple, uniformly bright, luminous disk source
models are becoming inadequate.  It is beyond the
scope of the present paper to model the
hydro-gravitational processes of quasar accretion,
but it is not unreasonable to speculate that such
regions of large turbulence dissipation rate
$\varepsilon$, high gas density $\rho$, and enormous
luminosity could rapidly produce a large number of
powerful point sources of light.  
Stars form in turbulent regions with mass $M_T
= (\varepsilon^6 \rho^{-5} G^{-9})^{1/4}$ in
times $\tau_G = (\rho G)^{-1/2}$, \cite{gib96}, with
$M_T$ limited only by radiation pressure differences that decrease as the
ambient luminosity increases
(the Eddington limit with small ambient luminosity
is
${M_E}_0 \approx 100 M_{\sun}$).  The luminosity of
such quasar superstars should increase rapidly with
$M_T \gg {M_E}_0$.  The Tully-Fisher relationship,
where spiral galaxy luminosity increases with
rotation rate, is another candidate for
enhanced turbulence causing an excess of
large stars with large luminosities. 

\section{Star-microlensing observations}
\label{secb}

The star-microlensing, or MACHO/EROS/OGLE
experiments, are operating in a very different
optical depth realm, and are therefore affected by
the detection statistics in profoundly different
ways. These experiments were justified by the
statistical inference of \cite{pac86}, later
confirmed by \cite{gri91}, that if the missing-mass
consists of stellar objects  the optical depth for
microlensing by a star in the Galaxy's Halo of a
star in the disk of the Large Magellanic cloud
would be
 $\approx (1/2) \times10^{-6}$. Thus, by observing
up to
$8 \times 10^6$ LMC stars nightly, the probability
of detecting a microlensing event would hopefully
 be above unity. Early reports of the detection of
such MACHO/EROS/OGLE events and claims of the
detection of the Halo missing-mass
 had to be corrected by \cite{sah94}, who
demonstrated that microlenses in the LMC halo were
about as important as microlenses in the Galaxy's
Halo, and the events would be indistinguishable to
available experiments. To date these experiments
have reported several LMC events (hundreds of
events toward the Galaxy bulge) but no detection of
the missing-mass, partly because of the ambiguity
about  whether the LMC microlenses are in the Halo
of the Galaxy or in the LMC itself, and partly, we
suggest, because the likelihood of MACHO clumping
and intermittency have been neglected for the
$10^{-6} M_{\sun}$ mass range.

The statistics of star-microlensing are profoundly
different than those for quasar-microlensing
because of the very  different optical depth
regimes. For star-microlensing, on any night, of
order 
$2 \times 10^6$ stars are sampled, and so there are
2\,000\,000 independent statistical tests asking
``is a solar-mass microlensing event underway.''
Since an event for solar-mass microlenses lasts
about 30 days, a new sample is available
monthly.  However, it is not an independent
sample since it involves the same objects only
slightly displaced by their relative velocities. 
The time required to assure complete temporal
independence of the sample is a few million years
and spatial independence of the sample would require
a different set of stars elsewhere on the sky. If
the missing-mass is not approximately solar but
rather terrestrial  (i.e.,
$10^{-6} M_{\sun}$), then events should be of
duration  only hours, and most would have escaped
detection in the main star-microlensing searches. 
However, some higher frequency searches for such
events have been  undertaken (\cite{alc98},
\cite{ren98}) and the claim  made that
non-detection constitutes proof that they do not
exist. No explanation is given for the fact that
rapid microlensing is observed in quasar
monitoring, \cite{sch96}, and we conclude that
star-microlensing is unable to detect and study the
missing matter of galaxies.  Why?

In \S \ref{sec2} we show that the failure of
star-microlensing to detect the missing Galactic
Halo matter is readily understood from the sparse
number of independent samples and because of the
expected non-Gaussian distributions of the PFP
masses of the microlensing objects within PGC clumps
as well as the likely non-Gaussian distributions of
the PGC clumps. Our purpose is to show that any
massive, primordial, population of small  condensed
objects would be expected to have entered a
nonlinear gravitational cascade of particle
aggregation, and the resulting highly intermittent
lognormal number density distribution would leave
too much uncertainty in the star-microlensing
statistics for any conclusions to be drawn about
the mean number density $n_p$ of
$10^{-6} M_{\sun}$ objects from limited
star-microlensing searches. Because
quasar-microlensing involves optical depths and
detection probabilities at least a factor of $10^6$
higher, it offers greater prospects for robust
sampling of mean  masses and mean densities of such
small dark-matter objects.
\cite{ker98} have also suggested inner-outer Halo
population differences might contribute to the
failure of star microlensing searches to find the
baryonic dark matter.

\section{Nonlinear hydro-gravitational condensation
of the primordial gas}
\label{sec1}

Star microlensing surveys have not aggressively
searched for planetoids as the missing Galactic mass
because the existence of such objects is impossible
by the generally accepted
\cite{jns02} theory of  gravitational instability.
However, Jeans's linear acoustic theory is subject
to vast errors, especially for very
strongly turbulent and very weakly turbulent flows
and for weakly collisional materials such as
neutrinos with enormous diffusivities $D$,
\cite{gib00a}bc.  Cold dark matter theories for
galaxy formation from nonbaryonic dark matter
condensations fail because the diffusivity $D$ for
such material is so large that the diffusive
Schwarz scale is larger than the corresponding
Jeans scale,
$L_{SD} \equiv ({D^2}/\rho G)^{1/4} \gg L_J \equiv
V_S / (\rho G)^{1/2}$, where $L_J$ values in CDM
theories are made small enough for galaxy size
CDM halos to form by assuming the nonbaryonic dark
matter is cold to reduce the sound speed
$V_S$.  However, no structures can form by
gravity on scales smaller than the largest Schwarz
scale according to
\cite{gib96}, and in this case the largest
Schwarz scale will be $L_{SD}$.  For a weakly
collisional nonbaryonic fluid, the criterion
$L_{SD} \gg L_J$ is equivalent to
${m_p}^{3/2}/\sigma \gg (kT\rho/G)^{1/2}$, where
$m_p$ is the particle mass, $\sigma$ is the
collision cross section, and $k$ is Boltzmann's
constant.  Substitution of $\rho$ and $T$ values
from the plasma epoch, when CDM halos were
presumably formed, give values about $10^{-12}$
$\rm kg^{3/2} m^{-2}$.   Typically assumed CDM
particle masses
${m_p} \ge 10^{-27}$ kg would require 
$\sigma$ values $\approx 10^{-28} \, \rm m^2$ for
the equality to fail, but realistic 
$\sigma$ values for CDM must be many orders of
magnitude less for its particles to escape
detection, $< \approx 10^{-43} \, \rm m^2$. 
Observed outer-halo galaxy scales of
$10^{22}$ m (324 kpc) suggest from
$L_{SD}$ that the nonbaryonic dark matter has
weakly collisional particles with $m_p$ values
closer to $10^{-35}$ kg, and this material
makes only a small ($\approx 1\%$) contribution to
the inner-halo galaxy density,
\cite{gib00b}.  

The case of strong turbulence was first
recognized by
\cite{cha51}, who suggested that a  ``turbulent
pressure''
$p_{T} = \rho V_{T}^2$ should simply be added to the
gas pressure $p$ in the Jeans length scale definition
$L_J
\equiv V_S/(\rho G)^{1/2}
\approx (p/\rho ^2 G)^{1/2}$, where $V_S \approx
(p/\rho)^{1/2}$ is the sound velocity, $p$ is
pressure, $\rho$ is density, and $G$ is Newton's
constant of gravitation.  This argument is flawed
for two reasons: firstly because pressure
without gradients offers no resistance to
gravitational forces (the basic flaw in Jean's
theory), and secondly because the turbulent
velocity is a function of length scale $L$. 
Substituting
$p_T$ for $p$ in the expression for $L_J$ rather
than adding and using the Kolmogorov expression
$V_{T}
\sim (\varepsilon L)^{1/3}$ gives the turbulent
Schwarz scale $L_{ST}
\equiv \varepsilon ^{1/2}/ (\rho G)^{3/4}$ of
\cite{gib96}, where $\varepsilon$ is the viscous
dissipation rate of the strong turbulence. If the
dissipation rate $\varepsilon \le \rho \nu G$,
where $\nu$ is the kinematic viscosity of the gas,
then the flow is non-turbulent at the scale of
gravitational domination and is controlled by
viscous forces at the viscous Schwarz scale
$L_{SV} \equiv (\gamma \nu/\rho G)^{1/2}$, where
$\gamma$ is the rate-of-strain of the flow,
\cite{gib96}. 
 
In gas
where $L_{ST} \gg L_J$, Jeans mass stars cannot
form because the turbulence forces overwhelm
gravitational forces at all scales smaller than
$L_{ST}$.  Such a situation may occur due to
supernova shock turbulence, in starburst regions and
in cold molecular clouds, and due to gravitational
infall turbulence near galaxy and PGC cores with
$\varepsilon \ge RT(\rho G)^{1/2}$ for strongly
turbulent gas regions on scales larger than
$L_{ST}$, where $V_S^2 \approx RT$ and $R$ is the
gas constant.  The case of interest in the present
paper has been overlooked by standard
cosmological models; that is, the hot gasses of the
early universe emerging from the plasma epoch with
weak or nonexistent turbulence, where both
$L_{SV}$ and
$L_{ST}
\ll L_J$.  

The physical significance of the Jeans scale $L_J$
is found by deriving the maximum length scale of
pressure equilibrium in a fluctuating density field
with self gravitational forces.  Pressure
equilibrium occurs if the gravitational free fall
time $\tau_G
\equiv (\rho G)^{-1/2}$ is greater than or equal to
the time required for acoustic propagation of the
resulting adjustment in the pressure field on scale
$L$; that is,
$\tau_G \ge \tau_P \equiv L/V_S$.  Solving gives $L
\le L_J$.  Thus $L_J$ is not a minimum scale of
gravitational instability as usually supposed, but
is a maximum length scale of acoustical pressure
equilibrium.  

This shows that there should be
fragmentation of the primordial gas at both the
Jeans scale $L_J
\approx 10^4 L_{ST}$ and the weakly turbulent or
nonturbulent Schwarz scales
$L_{ST} \approx L_{SV}$, \cite{gib96}.  From the
ideal gas law
$p = \rho R T$, the pressure can
adjust to density increases and decreases due to
gravitational condensations and void formations on
scales $L_J \ge L \ge L_{ST}$ keeping the
temperature nearly constant without radiant heat
transfer.  Voids forming at scales larger than
$L_J$ cannot adjust their pressure fast enough to
maintain constant temperature, causing a temperature
decrease within the voids.  Radiant heating of such
voids larger than
$L_J$ by the surrounding warmer gas therefore
accelerates their formation in the expanding
proto-galaxy, causing fragmentation of the
primordial gas into proto-globular-cluster (PGC)
blobs with mass of order $10^{36}$ kg, or
$10^{6} M_{\sun}$.  Since this instability has
nothing to do with the Jeans theory, the scale
$L_{IC} \equiv (RT/\rho G)^{1/2}$ has been termed
the initial condensation scale, \cite{gs99a}b.

Because pressure forces propagate by
electromagnetic radiation in plasmas the
sound speed in the plasma epoch is nearly
the light speed $c$, giving Jeans scales $L_J$
larger than the Hubble scale of causal connection 
$L_H \equiv ct$ and therefore no possibility of
structure formation in the baryonic component of
matter by the Jeans theory.  However, we have seen
that the Jeans scale is generally not the criterion
for gravitational structure formation, but instead
gives the maximum scale of pressure equilibrium by
acoustic propagation.  Thus we can conclude that
the small inhomogeneities in the CMB observed by
COBE and other experiments are not acoustic as
often assumed, but are more likely remnants of
the first formation of structure due to gravity. 
The power spectrum of COBE temperature fluctuations
peaks at a length scale ${L_{H}}_{FS} \ll L_H$ much
smaller than the horizon, or Hubble, scale $L_H$ at
the time of photon decoupling, suggesting the first
structure formation must have occurred previously
to match the spectral peak scale to a smaller
horizon scale
${L_{H}}_{FS}$ existing at some earlier time of
first structure,
\cite{gib00b}.

Formation of proto-super-cluster to
proto-galaxy mass fragments from void formations
was possible in the plasma epoch, but with slight
increases in density from gravitational
condensation because the universe age $t$ was at
all times less than the gravitational free fall
time $\tau_G
\equiv (\rho G)^{-1/2}$ and because of non-baryonic
diffusive compensation,
\cite{gs99b}. Expansion of the universe enhances
formation of voids and retards condensation of
fragments, but amplitudes of both are damped
by rapid diffusive density compensation as
the non-baryonic component
fills the minima and exits the maxima formed in the
baryonic component. Fragmentation and fractionation
of the primordial plasma was first triggered when
the Hubble scale of causal connection $L_H \equiv
ct$  decreased to scales less than the viscous or
weakly turbulent Schwarz length scales
$L_{SV} \approx L_{ST}$,
\cite{gib96}.  Remarkably, the baryonic matter
density $\rho(t)$ from Einstein's equation at
$t = 10^{12}$ s of $10^{-17}$ kg $\rm m^{-3}$,
\cite{win72},
matches the density of globular star clusters
(GCs).  We take this $\rho$ value to be that of the
primordial gas at $t = 10^{13}$ s, and further
evidence of structure formation beginning at about 
$t = 10^{12}$ s.  Observations of $\rho$
$\approx 10^{-17}$ kg $\rm m^{-3}$ in dim and
luminous globular clusters suggests the
internal evolution of PGCs has been quite gentle and
nondisruptive so that dark PGCs with PFPs in
less advanced stages of their accretional
cascades may be considered metastable, with
average densities $\rho_{PGC} \approx \rho_{IC}$ and
masses $M_{PGC} \approx M_{IC}$ that depart slowly
from these initial values.  

Additional evidence
suggesting the time of first structure was about
$t_{FS}
\approx 10^{12}$ s, or
$30 \, 000$ years, is that this is the time when the
supercluster mass of $10^{46}$ kg matches the Hubble
mass $\rho(t) (ct)^3$, \cite{gib97a}. Taking the
horizon scale at this time equal to the viscous
Schwarz scale
$L_{SV}
\equiv (\gamma \nu / \rho G)^{1/2}$ with $\gamma
\approx t^{-1}$ gives a kinematic viscosity $\nu$ 
of $5 \times 10^{26}$ $\rm m^2$ $\rm s^{-1}$,  which
matches the photon viscosity $\nu \equiv l_C c$
obtained from the mean free path for Thomson
scattering $l_C = 1/n_e  \sigma_T$ with free
electrons of the plasma times the speed of light
$c$, where $n_e$ is the free electron number
density and $\sigma_T$ is the Thomson scattering
cross section,
\cite{gib99b}.  This enormous viscosity (a million
times that of the Earth's upper mantle) gives a
horizon scale Reynolds number $\rm Re \equiv c^2
t/\nu \approx 200$ which is slightly above
critical, consistent with evidence from COBE of
weak turbulence in the primordial plasma, and
precludes any interpretation of the observed CMB
temperature anisotropies as ``sonic peaks'' since
any such powerful plasma sound would be promptly
damped by viscosity, even if its unknown source
could be identified, \cite{gib00b}.

Evidence of proto-supercluster to proto-galaxy
formation in the plasma epoch can also be inferred
from the advanced morphology of superclusters, with
$10^{24}$ m voids bounded by ``great walls'' as
shown by deep galaxy maps, and by galaxy rotation
rates of order $10^{-15}$ rad
$\rm s^{-1}$ consistent from angular momentum
conservation as the galaxies expand with our
estimated time of proto-galaxy fragmentation at
$\approx 10^{13}$ s, with angular velocity $\Omega$
and strain rate
$\gamma$ both near  $10^{-13}$ rad
$\rm s^{-1}$. 
Thus, we assume that
the neutral primordial gas emerging from the plasma
epoch was fragmented into
$\approx 10^{41}$ kg proto-galaxy blobs of hydrogen
and helium with a very uniform density $\rho
\approx 10^{-17}$ kg
$\rm m^{-3}$ reflecting the time $t_{FS} \approx
10^{12}$ s of first fragmentation and
proto-superclusters, and
$\Omega \approx 10^{-13}$ rad
$\rm s^{-1}$.

Turbulent mixing scrambles density fluctuations from
large scales to small to produce nonacoustic
density nuclei that move with the fluid, and these
trigger gravitational condensation on maxima and
void formation at minima.  Condensation on acoustic
density nuclei; that is, density maxima moving with
the sound speed as envisaged by Jeans, is not a
realistic description of gravitational structure
formation under any circumstances, as shown by
\cite{gib96}.  The smallest scale
$\rho$ fluctuation produced by turbulent mixing is
at the Batchelor length scale 
$L_B \equiv (D/\gamma)^{1/2}$, \cite{gib91},
corresponding to a local equilibrium between
convective enhancement of
$\nabla \rho$ by the rate-of-strain $\gamma$  and
smoothing of $\nabla \rho $ by the molecular
diffusivity $D$,  near points of maximum and
minimum $\rho$ moving with the fluid velocity $\bf
\vec v$, \cite{gib68} and \cite{gs99a}. 

The Jeans theory fails because it relies on the
Euler momentum equations that neglect viscous
forces, and because it employs linear
perturbation stability analysis, which neglects the
nonlinear inertial-vortex forces that dominate
turbulent flows, \cite{gib99a}. Viscous forces are
necessary to stabilize the rapidly expanding flow of
the early universe for times $t$ before structures
form, with rate-of-strain $\gamma$ values of order
$t^{-1}$,
\cite{gib99b}.  Expanding inviscid flows are
absolutely unstable (\cite{lan59}, \cite{gib00b}),
so a major result of the COsmic Background
Experiment (COBE) was a demonstration that
temperature fluctuations
$\delta T/T$ were $\le 10^{-5}$ rather than
$\ge 10^{-2}$ to be expected if plasma
epoch flows before transition to neutral gas were
fully turbulent.  In the absence of structure to
produce buoyancy forces, only viscous
forces (small Reynolds numbers) can explain the
absence of turbulence that is manifest in the
homogeneity of the cosmic microwave background
(CMB) temperature fluctuations.  

Linearity and the Jeans theory cannot be salvaged
by arguing that
$\delta \rho / \rho$ is small so that very long
times would be required for the gravitational
structure formation to ``go nonlinear''.  The time
required to form structure by gravity is
$\approx \tau_G \equiv (\rho G)^{-1/2}$ where
$\tau_G$ is the gravitational free fall time, and
nearly independent of the magnitude of the dominant
density fluctuations on scales larger than the
largest Schwarz scale.  The mass of such a
fluctuation $M'(t) \approx M'(0) exp[2 \pi
(t/\tau_G)^2]$,
\cite{gib99b}.  Thus, the
$10^{-3}$ difference between the observed CMB
density fluctuations and the larger values expected
for turbulence only causes a 45\% increase in
the time required for $M'(t)$ to reach planetoidal
values.  For $\rho = 10^{-17}$ kg $\rm m^{-3}$,
$\tau_G = 3.9 \times 10^{13}$ s or $1.2 \times
10^6$ years.  Starting from $10^{-17}$ kg
$\rm m^{-3}$, the time  to reach density
$\rho = 10^{0}$ kg $\rm m^{-3}$ is $3.1
\times 10^6$ years ($10^{14}$ s) and $< 10\%$ more
to reach 
$10^{3}$ kg
$\rm m^{-3}$ and quasi-hydrostatic equilibrium. 
This first condensation process of the universe has
a profound effect on everything else that happens
afterwards.  The dynamics shifts from that of a
collisional gas to that of weakly collisional
warming gas blobs in an increasingly viscous,
cooling, gaseous medium, evolving slowly toward a
cold dark metastable state of no motion as the
baryonic dark matter of galaxies, as observed by
\cite{sch96}. The sites with the highest likelihood
of instability to form the first stars and quasars
are the cores of PGCs and the cores of PGs, at this
same time $ t \approx 10^{14}$ s with $z \approx  
264$.  A large population of quasars at extremely
high redshifts
$z$ is indicated by dim point sources of hard
X-rays detected by the Chandra satellite,
\cite{mus00}.  A survey of star-forming galaxies
for $3.8 \le z \le 4.5$ after improved dust
extinction corrections shows no indication of a
peak in the (star formation rate)
SFR/comoving-volume and therefore a monotonic
increase in SFR/proper-volume with increasing $z$
toward a peak at some high $z \ge 4.5$ value,
\cite{sti99},
presumably also at $z \approx  
264$, compared to a peak at $z = 1 \-- 2$ assumed by
many authors following the ``collapse'' of gas into
CDM halos to form galaxies at $z \approx 5$ for the
standard CDM model.  Galaxies never
collapse after their formation at $z \approx 1000$
in the present scenario.  Clusters of galaxies
observed at z = 3.09, \cite{sti00}, also present no
difficulties for our scenario of first structure
growth.

The kinematic viscosity of the neutral primordial
gas formed from the cooling plasma at decoupling is
dramatically less than that of the plasma, with
$\nu \approx 5
\times 10^{12}$ $\rm m^2$ $\rm s^{-1}$,
\cite{gib99b}.  Thus the viscous Schwarz scale
$L_{SV} = (5 \times 10^{12} \times 10^{-13} /
10^{-17} \times 6.7
\times 10^{-11})^{1/2}$ = $2.7 \times  10^{13}$ m,
and the viscous Schwarz mass $M_{SV} \equiv
L_{SV}^3 \rho = (2.7 \times  10^{13})^3 \times
10^{-17}$ is
$2
\times 10^{23}$ kg, or
$10^{-7} M_{\sun}$, which we take to be the
minimum expected mass of a primordial fog particle.
The minimum PFP mass increases to $\approx 5 \times
10^{24}$ kg if fossil turbulence effects are
included, taking $\gamma_o = 10^{-12}$ $\rm s^{-1}$
as a fossil vorticity turbulence remnant of the
time of first structure $10^{12}$ s,
\cite{gib00b}c. The weak turbulence levels possible
from the COBE observations increase the condensation
mass to  maximum $M_{ST} \le 10^{25}$ kg values
that are still in the small-planetary-mass range. 
The formation of these gaseous planetoids represents
the first true condensation by gravity since the
Big Bang, where the mass
density $\rho$ increases with time.  Whatever
small turbulence may have existed at decoupling
will now be damped by buoyancy forces as the PFPs
begin their gradual evolution from hot gas blobs to
freezing cold liquid-solid Neptunes at $z \approx
3$, with a gentle, complex, accretional cascade to
form stars that results in practical
invisibility of these PFPs to star-microlensing
searches.  Frozen hydrogenous planetoids are stable
to evaporation at the temperature of the present
universe for masses larger than about
$10^{-8} M_{\sun}$, \cite{der92}.  No account has
been made for possible increases in PFP atmosphere
size or increased PFP number density on their
fraction of supernova energy absorbed with
increasing
$z$, even though
$\Lambda \ne 0$ and ``dark energy'' inferences
depend on only 30\% dimming of
supernova Type Ia events at the larger $z$ values to
justify claims of
accelerations in the universe expansion.  

\section{Intermittent lognormal  number density of
hydrogenous, primordial, small-planetary-mass
objects}
\label{sec2}

Star-microlensing estimates (\cite{alc98}) of the
contribution  to the missing-mass of the Galaxy
Halo by compact Halo objects in various mass ranges
 have assumed that the objects are uniformly
distributed in space.  This assumption is highly
questionable, especially for objects in the 
small-planetary-mass range 
$M_p \approx 10^{-6} M_{\sun}$. If such objects are
sufficiently compact to cause microlensing,  and
sufficiently numerous to dominate the Galaxy mass,
they must be primordial, and consist of H-He in
primordial proportions.  They comprise the initial
material of construction for the first small stars
in PGCs, and an important stage in the process of
such star formation.   Non-baryonic materials are
too diffusive to become compact, and no cosmological
model has produced this much baryonic mass that is
not H-He.  
  
Starting from a nearly uniform distribution of
$10^{6} M_{\sun}$ PGC-mass clumps of hot-gas PFPs in
the proto-galaxy (PG), all of the PFPs must
have participated  to some extent in nonlinear,
gravitational, accretion cascades culminating for
luminous GCs in the formation of the first small
stars simultaneously, giving an abrupt end to
the ``dark ages'' of the universe at $z
\approx 264$, versus $5$ much later as usually
assumed, with the first large stars forming
simultaneously with reevaporation and the first
strong turbulence at PGC-cores near PG-cores as
Population III super-stars, with their supernovas,
black holes, and proto-quasars, and probably one or
less for each proto-galaxy so that the universe
never reionized as often assumed. Those PGCs in the
outer regions of PGs should cool more rapidly and
be less successful in accretion cascades of their
PFPs than those nearer PG-cores. Pairs of PFPs
should form first, and then pairs of pairs, etc.,
producing a complex array of nested clumps within
each PGC, and large voids in between. Large drag
forces from the fairly dense, weakly turbulent,
inter-PFP gas should produce small PFP terminal
velocities from its drag compared to virial
velocities, and the large number of neighboring
PFPs and their clumps should tend to randomize PFP
velocity directions.

Such an accretion
process  produces a highly non-uniform PFP number
density $n_p$, with  the potential for
large undersampling errors in star-microlensing
estimates of the mean number density $\langle n_p
\rangle$ of such objects in the Halo
from the resulting small numbers of independent
samples of $n_p$.  A sample consists of
several observations of each star for times longer
than the expected microlensing time for a number of
stars sufficient to give at least one event, where
the sampling time required depends on how many
objects are expected and whether the Galaxy Halo
consists of a uniform distribution of objects in
the sampled mass range or a nonuniform
distribution.  It takes longer to acquire a sample
if the probability density function (pdf) of $n_p$
is nonuniform and much longer if it is extremely
intermittent.  An independent sample is one taken
from stars separated from other samples by
distances larger than the largest size of clumps of
such objects or any clumps of such clumps.  If the
same stars are used for each sample, one must allow
time between samples so that the sampled field of
objects can move away and be replaced by an
independent field.  To avoid undersampling error,
one must account for nonuniformity of the objects
sampled and take sufficient numbers of independent
samples to achieve an acceptable level of
statistical uncertainty in the estimated mean
$\langle n_p
\rangle$.

\subsubsection{Clustering models for the accretion
of primordial-fog-particles (PFPs)}

Detailed modeling of the accretion of PFP objects
within  the proto-globular-cluster (PGC) initial
condensation masses is an important topic, but is
rather complex and therefore beyond the scope of the
present paper.  Rates of PFP merging depend on
relative velocities, frictional damping, and
degrees of condensation of the PFP gaseous
atmospheres. A gas epoch existed before
$z = 3$ with no possibility for condensation of the
primordial gas to liquid or solid states because
universe temperatures were above hydrogen and
helium dew point and freezing point temperatures. 
Thus rates of merging of the large, entirely
gaseous PFPs existing then should have been more
rapid than the colder, more compact PFPs to appear
later with condensation to liquid and solid states
in the colder universe.  Frictional and tidal
forces between gassy PFPs were significant,
reducing velocities and merging rates and
permitting radiative cooling and further
condensation toward more collisionless PFP states. 
Principles of collisionless dynamics 
(\cite{bin87}) describing star interactions often
may not apply to PFP interactions, particularly in
the early stages of PFP merging when large scale
gas atmospheres about the objects existed.  Neither
must stellar virial theorems apply to PFPs  because
frictional forces of PFP extended gaseous
atmospheres will reduce their relative velocities 
below virial values if they have been exposed to
radiation, tidal, or convective heating flows that
can cause re-evaporation.  

As an exercise, we can compare star clustering with
PFP clustering.  It is well known  from stellar
dynamics that the third
star in a triple must lie at least three times the
radius of a binary away from the center.  An
accretion cascade of triples to form triples of
triples, etc., within a uniform cluster of stars,
would result in an ever decreasing  average mass
density of the star cluster, rather than a constant
or increasing density as expected for frictionally
interacting PFPs within a
proto-globular-cluster (PGC) as they accrete in a
bottom-up cascade to form globular cluster stars. 
If there were $n$ stages in a tripling cascade and
we assumed the third member of the $k$-th triple
cluster must be four times the radius $R_k$
separating the associated binary of
$(k-1)$-th clusters, where $1 \le k \le n$, then
the mass of the $n$-th cluster is 
$M_n = 3^n M_{star}$, and the radius $R_n = (4m)^n
d$, with star size $d$ and binary star separation
$R_{bs} = md, m \gg 1$.  This would result in a
monotonic decrease in the mass density of star
clusters $M_n = 3^n M_{star}/(4m)^{3n} {R_{bs}}^3$ 
with increasing $n$.  If the initial mass density
of unclustered  stars is 
$\rho_{0} = M_{star}/(4md)^3 \approx
\rho_{star}/(4m)^3$, where
$\rho_{star}$ is the star density, then $\rho_n
/\rho_{star}  = 3^n (\rho_0/\rho_{star})^{3n}$,
which also decreases rapidly with increasing values
of $n$ since $\rho_0/\rho_{star} \ll 1$.  

Such a star tripling cascade is possible, but is
slow.  Times between collisions $t_{col} =
1/(\sigma_{star} v n)$ are large.  For example, in
a virialized globular cluster with $\sigma_{star}
\approx 10^{18}$
$\rm m^2$,
$v \approx 8 \times 10^3$ $\rm m \> s^{-1}$, and 
$n \approx 10^{-51}$ $\rm m^{-3}$ we find $t_{col}
\approx 10^{29}$ s, which is $\approx 3 \times
10^{11}$ larger than the age of the universe.  On
the other hand, times $t_{PFP}$ for collisions of
PFPs 
$t_{PFP} \approx
L_{sep}(\rho_{PFP}/\rho_{0})^{2/3}/v$  are much
shorter, where $L_{sep} = (M_{PFP}/\rho_{0})^{1/3}$
is the PFP separation distance, $\rho_{0}$ is the
mass density of the PGC,
$v$ is the relative velocity, and $\rho_{PFP}$ is
the PFP density. For the same relative velocity  $v
\approx 8 \times 10^3$ $\rm m \> s^{-1}$,  for
$\rho_{PFP}/\rho_{0} = 10$, 
$\rho_{0} = 10^{-17}$ $\rm kg \, m^{-3}$, and 
$L_{sep} = 2 \times 10^{13}$ m, we find $t_{PFP}
\approx  10^{10}$ s, which is  much less than the
age of the universe of $10^{13}$ s when PFPs were
first formed.  Relative velocity values for PFPs
then were certainly much less than virial values
$(2GM/R)^{1/2}$, with collision times as much as a
few times the universe age as a lower bound based
on $ v \approx \gamma R$ from the universe
expansion value $\gamma = 1/t$.    

As another exercise, consider the commonly assumed
virial theorem for collisionless objects such as
stars.  The virial theorem applied to PFPs within a
PGC soon after fragmentation
would incorrectly equate the kinetic energy per unit
mass $v^2/2$ of each PFP to its potential energy per
unit mass  $GM/R$ within the PGC of mass $M$ and
radius
$R$.  By Hegge's law (\cite{bin87}), the outer
orbital velocity of stars, from the  application of
the virial theorem to a star cluster,  must exceed
the dispersion velocity of the stars in the
cluster, or else any internal hierarchical star
structures will be destroyed.  Hegge's law is
appropriate to star clusters because stars are
virtually collisionless.  However, frictional and
tidal forces between interacting PFPs prevent
virialization and enhance merging.  For the
purposes of the present paper, it suffices to
postulate that the aggregation  cascade at every
stage from PFP to star mass within a PGC is
nonlinear and approximately self-similar, leading
to clumping and  increased intermittency of the PFP
number density as the number of stages increases.

It is not known what fraction of the gas of a
proto-galaxy fragments into proto-globular-clusters
simultaneous with PGC fragmentation into  PFPs. 
Since there are only about 200 globular clusters
(GCs) in the Milky Way Halo, and this number is
typical for nearby galaxies, one might think that
PGC formation were a rare event since about a
million PGCs are required to equal the mass of a
proto-galaxy and only 0.02\% are observed in our
nearby galaxies close enough to sample.  A more
likely scenario is that $\approx 10^{6}$ PGCs formed
per galaxy as expected, but most PGCs have remained
dark with so few stars that they have escaped
detection, either in the bright, dusty, galaxy core
where the clustering of stars is more difficult to
discern, or in the dark halo where external
triggers of star formation are rare and most PGCs
are still only partially evolved.  Some
fraction of the original PGCs of a galaxy will
be disrupted to form the interstellar medium of
the disk and core where rogue PFPs
will generally dominate the ISM mass.  Dynamical
constraints on dark clusters of objects such as
PGCs are examined by
\cite{car99}, and our value of $[M_C / M_{\sun} ,
R_C /pc ] = [10^6 , 8]$ is well within their range
of values not excluded. 

An authoritative 
book on globular cluster systems suggests
that observations of these systems have
``outstripped theoretical work'' on how the systems
and the globular clusters themselves are formed,
especially young globular clusters, \cite{ash98}. 
The accumulation of evidence supports early
suggestions of these authors that galaxy
merger-induced starbursts are favorable
environments for globular cluster formation, as
dramatically confirmed in the Hubble Space
Telescope (HST) images of the merger galaxy
NGC 3256 where more than 1000 bright blue objects
have been identified as young globular clusters,
\cite{zep99}.  Our suggestion is that the bright
system of young GCs observed within $10^{20}$ m of
the central region of the NGC 3256 merger can be
readily explained as a small fraction of the many
dark, primordial, PGC-PFP systems available in the
merging galaxies that have been brought out of cold
storage by tidal forces of the merger process
accelerating the accretion of PFPs to form stars. 

With the recent high resolution and infrared
capabilities of the HST, many dense star clusters
near the core of the Milky Way have been revealed. 
These clusters have masses of order
$10^6 M_{\sun}$ expected for PGC objects formed at
the initial condensation $L_{IC}$ scale at
decoupling.  Densities up to
$10^{-13}$ kg $\rm m^{-3}$ are indicated, with
massive $100 M_{\sun}$ stars
 reflecting the large $L_{ST}$ scales caused by the
high turbulence levels from the frequent supernovas
and powerful radiation and winds.  Supernovas,
radiation and winds are also mechanisms for
increasing PGC density for PGCs close to a galaxy
core because they increase friction by producing
higher gas densities in the inter-PFP medium within
evolving PGCs.   

Super-star clusters with mass $10^6 M_{\sun}$  and
$\rho \approx 10^{-15}$ kg $\rm m^{-3}$  are
reported by \cite{oco94} in the core of the dwarf
galaxy NGC 1569, observed from the HST even before
its repair, confirming the 1985 claim of Arp and
Sandage that these objects were not stars but
super-star-clusters. 
\cite{mar97} conclude that the brightest of the
several super-star-clusters NGC 1569A in the NGC
1569 galaxy is actually a superposition of two
clusters separated by a distance close to their
sizes, so the clusters are apparently in orbit. 
They are identified as ``young globular star
clusters''.  Similar young globular star clusters
are reported by
\cite{ho96} at the center of NGC 1705, an amorphous
galaxy,  by \cite{hol96} in the interacting
galaxy 1275, and by \cite{wat96} in the
starburst galaxy NGC 253.  In the two colliding
galaxies NGC 4038/4039 termed The Antennae, a
population of over 700 blue pointlike objects were
identified by
\cite{whi95} as young globular clusters formed as a
result of the merger, and are therefore possibly
dark PGCs brought out of cold storage. The size of
the objects is given as 18 pc, giving a density of 
$10^{-17}$ kg $\rm m^{-3}$ that is very close to
the density at the time of PGC formation from
primordial gas, and represents a typical density of
Halo GCs.  Thus we have evidence that dark
proto-globular-clusters (DPGCs) exist in abundance
in a variety of galaxy types.  If DPGCs consist of
PFPs, no significant fraction of their original
number of PFPs have apparently evaporated by
collisionless dynamics mechanisms.  Instead, the GC
mass-density $\rho$ changes observed are to larger
rather than smaller values.  The original
condensation mass
$10^6 M_{\sun}$ has not changed.

For PGCs formed in the outer regions of
proto-galaxies (PGs) that were most likely to have
maximum turbulence and minimum gas density and thus
maximum PFP mass and separation and most rapid
cooling to a collisionless state, a
more stellar-like dynamics of clustering might seem
appropriate with consequent decrease in the
average PGC mass density.  However, we find no
evidence that GC densities outside galaxy core
regions ($\approx 10^{20}$ m) deviate appreciably
from
$\rho \approx 10^{-17}$ kg $\rm m^{-3}$, suggesting
that PGCs are gas stabilized by the possibility
of reevaporation of their PFPs. The PGCs themselves
presumably dispersed, along with the luminous GCs,
from locations within an initial maximum galaxy
diameter of
$10^{20}$ m to the present size of
$10^{21}$ m, partly due to the expansion of the
universe and perhaps partly from their more
collisionless dynamics compared to those near 
galaxy cores.  Initial Hubble flow velocity
differences at PG scales
after decoupling were $\delta v_r =
\gamma d = 10^{-13} \times 10^{20} = 10^7$ m $\rm
s^{-1}$, giving rapid expansion of the galactic
scale protovoids.  The average mass density of
galaxies is certainly substantially less today, with
$\rho \approx 10^{-21}$ kg $\rm m^{-3}$, than the
average mass density of $\rho \approx 10^{-17}$ kg
$\rm m^{-3}$ for the PGs discussed in \S
\ref{sec1}.  

Evidence for the existence of Halo dark matter in
the form of cold $\rm H_2$ gas clouds clumped in
dark clusters is claimed from measurements of
diffuse $\gamma$-ray flux,
\cite{dep99}.  Similarly, cold self-gravitating
hydrogen clouds explain ``extreme
scattering events'' of compact radio quasars,
\cite{wal98}.  Both of these
gas-cloud interpretations are consistent with and
support our postulated dark metastable PGC-PFPs as
the inner-galaxy-halo dark-matter, with their
inter-PFP-gas and PFP-atmospheres, except that we
require that most of the PGC mass be in the form of
mostly-frozen solid-liquid PFPs as the source of
this gas that resists changes in the average PGC
mass density by shifts in the equilibrium gas
density.  For cold
$\rm H_2$ clouds to exist completely as gas in
clumps with PGC densities and sizes, huge turbulence
dissipation rates of
$\varepsilon
\ge 3.4
\times 10^{-6}$ $\rm m^2$ $\rm s^{-3}$ would be
required to prevent their condensation to form
stars at $L_{ST}$ scales, compared to primordial
$\varepsilon$ values of only $\approx 10^{-13}$
$\rm m^2$ $\rm s^{-3}$ estimated at the plasma-gas
transition from the measured CMB temperature
fluctuations leading to PFPs,
\cite{gib99b}.  No source of kinetic energy is
available to produce or sustain such turbulence, and
without such a source to prevent
gravitational condensation the turbulence would
dissipate in a few million years and the PGC mass
gas cloud would collapse in a powerful star-burst
of super-star formation.

\subsubsection{Intermittent lognormal $n_p$
probability density function}

The average separation distance $L_p$ between 
small-planetary-objects (PFPs) for a galaxy at
present is about
$ (M_p/\rho_{halo})^{1/3} \approx 7\times 10^{14}$
m for a galactic density
$\rho \approx 10^{-21}$ $\rm kg \> m^{-3}$, with
object size 
$r_p \approx (M_p/\rho_p)^{1/3} \approx 4\times
10^{6}$ m for a density $\rho_p \approx 3 \times
10^{3}$ $\rm kg \> m^{-3}$ corresponding to
condensed H-He at halo temperatures, giving
approximately 8 decades for the cascade  from
separation to accretion scales of $M_p$ mass
objects. The number density
$n_p (r)$ of such planetary objects per  unit
volume becomes an increasingly intermittent random
variable (rv) as the averaging volume size $r$
decreases.   If we assume most of the nonlinear
gravitational cascade  is self-similar, then the
probability density function of $n_p (r)$ will be
lognormal with an intermittency factor $I_p (R/r)
\equiv \sigma ^{2}_{ln [n_p (r) / n_p (R)] }$,
where $\sigma ^2$ is the variance about the mean,
that increases as the range of the cascade $R/r$
increases as 
\begin{equation}
\sigma ^{2}_{ln [ n_p (r) / n_p (R)] } \approx
\mu_p ln (R/r)
\label{eqz}
\end{equation} where $R$ is the largest scale of
the cascade, $r_p \le r$ is the smallest, and
$\mu_p$ is a universal constant of order one.  The
expression is derived by expressing $n_p (r) / n_p
(R)$ as the product of a large number of breakdown
coefficients $b_m = n_p (k^m r) / n_p (k^{m+1}r)$
for  the $0 \le m \le l$ intermediate stages of
the  cascade with scale ratio $k$ sufficiently
large for the random variables $b_m$ to be
independent.   Taking the natural logarithm of the
product
\begin{equation} {n_p (r) \over n_p (R)} = {n_p (r)
\over n_p (kr)} {n_p (kr) \over n_p (k^2 r)} {n_p
(k^2 r) \over n_p (k^3 r)} ... \\ {n_p (k^{l-1}r)
\over n_p (R)} = b_1 b_2 ... b_l
\end{equation} gives
\begin{equation} ln [{n_p (r) / n_p (R)}] = ln
(b_1) + ln (b_2) + ... + ln (b_l).
\end{equation} By the  central limit theorem, the
random variable 
$ln [{n_p (r) / n_p (R)}]$ is Gaussian, or normal, 
because it is the sum of many identically
distributed, independent random variables 
$ln (b_m)$, which makes the random variable
$[{n_p (r) / n_p (R)}]$ lognormal.  The variance of
a normal rv formed from a sum of rvs is the sum of
the variances of its components. Since there are
$l$  breakdown coefficients $b_m$, and $R/r = k^l$,
then
\begin{equation}
\sigma ^{2}_{ln [ n_p (r) / n_p (R)] } 
\approx l \times \sigma ^{2}_{ln (b_m) } =  {ln
(R/r) \over ln (k)} \sigma ^{2}_{ln (b_m) } 
\end{equation} which gives Eq. (\ref{eqz}) and its
constant
\begin{equation}
\mu_p \approx {\sigma ^{2}_{ln (b_m) } \over ln
(k)}.
\end{equation} The universal constant $\mu_p$ is
always  positive because $\sigma ^{2}_{ln (b_m) }$
is positive and $k > 1$.

If the intermittency factor $I_p$ is  large, then
estimation of the mean number density
$\langle n_p \rangle = n_p (R)$ from a small number
of samples becomes difficult.  A single sample
gives an estimate of the mode, or most probable
value, of a distribution, and the mean to mode
ratio $G$ (the Gurvich number) for a lognormal
$n_p$ distribution is 
\begin{equation} G \equiv 
\frac{{\langle n_p \rangle}_{mean}}{{\langle n_p
\rangle}_{mode}}  = exp (3I_p/2).
\label{eqa}
\end{equation}
  Taking $\mu_p \approx 1/2$, 
\cite{ber94}, and substituting $R/r_p = 7 \times
10^{14} / 4 \times 10^{6}$ gives $I_p = 9.5$. 
Thus, $G = 1.5 \times 10^6$  from Eq. (\ref{eqa}),
which is the probable undersampling error for
$\langle n_p \rangle$  from a single sample.
Increasing $M_p$ has no effect on the intermittency
factor $I_p$ or the undersampling error $G$ as long
as the aggregation cascade is complete since the
ratio
$R/r_p$ is unaffected by $M_p$.

The intermittent lognormal distribution function
for $n_p (r)$ inferred above is generic for
nonlinear cascades over a wide range of scales and
masses and with a wide range of physical
mechanisms, since it is based on the central limit
theorem which gives nearly Gaussian distributions
even when the summed component random variables are
not identically distributed and are not perfectly
independent as long as there are many components.   

\subsubsection{Implications for star-microlensing
versus quasar-microlensing studies}

The EROS and MACHO collaborations  (\cite{alc97},
\cite{ren98}, \cite{alc98})  estimate they should
have encountered about 20 star-microlensing events
in their two years of observing with time periods
corresponding to 
$M_p \approx 10^{-6} M_{\sun}$, assuming such
objects are uniformly distributed in the Halo with
$\langle n_p \rangle R^3 M_p \approx M_{Halo}$.
Because they encountered none, they conclude that
the contribution of $M_p$ objects  can be no more
than 1/20 of $M_{Halo}$.  However, since $n_p$ is
expected to be a lognormal random variable (LNrv)
with intermittency factor $I_p = 9.5$, the most
that can be said from this information is that the
modal mass $\langle n_p \rangle _{mode} R^3 M_p$ of
such objects is less than  $M_{Halo}/20$. Since the
mean mass 
$\langle n_p \rangle R^3 M_p 
\approx G \langle n_p \rangle _{mode} R^3 M_p \le
M_{Halo} G/20 $, this is no constraint at all on
$M_p \approx 10^{-6} M_{\sun}$  as the mass of Halo
objects dominating
$M_{Halo}$ because $G \approx 10^{6} \gg 20$. 
About 
${2G}/{20}$ y of observing time, with no
detections, would be needed  to make the claimed
exclusion, or 150,000 years.
Furthermore, 
150,000 years assumes all LMC stars are seen
through PGCs even though these cover a small
fraction of the sky and the PGC density itself is
likely to be an intermittent lognormal.

As mentioned, another serious problem with the
EROS/MACHO sampling method is a lack of
independence of the samples.  Both collaborations
are constrained because  the solid angles occupied
by the stars of the Large Magellanic Cloud are
smaller than those expected for dark
proto-globular-clusters (PGCs) of PFPs.  Assuming a
PGC diameter of $4.6 \times 10^{17}$ m at a
distance of $10^{21}$ m, each PGC occupies a
fraction $1.7 \times 10^{-8}$ of the sky, or 1.7\%
for a million PGCs per galaxy.  Therefore, most
MACHO/EROS stars are likely to be in areas between
PGCs, so they probably will see nothing, as they
have done.   In order to give independent samples
of the Halo mass, stars chosen for
star-microlensing tests must be separated by
angular distances larger than the separation 
between the largest clumps.  The time for a dark
PGC to drift its diameter in the Halo is about
$10^{12}$ s.    Three hundred independent samples
are required to achieve 50\% accuracy with 95\%
confidence in estimating the expected value of a
LNrv with $I_p = 9.5$ (\cite{bak87}), requiring
$\approx 10^7$ y of observations toward the LMC to
achieve this accuracy.  This is very impractical.

A million PGCs in
the Galaxy Halo should occupy about 2\% of the
sky, so if they were uniformly distributed only 2\%
of the LMC stars would intersect a PGC. This in
itself would justify a claim that PFPs fail as the
missing mass based on the
assumption (questioned herein)
that particle density
$n_p$ distribution function is uniform.  We have
shown that even if all the LMC stars were seen
through PGCs, the indicated 
$\langle n_p
\rangle$ would still be underestimated since the
Gurvich factor for any PGC is likely to be much
larger than 200  ($G \approx 10^6$).  Quasar
microlensing is much less affected by
intermittency effects since as shown in \S
\ref{seca} the number of microlensing masses seen
projected in front of the quasar's luminous disk is
likely to be of order
$10^{8}$. Because no model as yet exists for the rapid
brightness fluctuations observed, we do not yet
know the role of structure in the quasar accretion
disk and the distribution of implied masses.

\section{Conclusions}
\label{secc}

Quasar-microlensing signals have continuous 
fluctuations at short time scales indicating the
lensing galaxy mass is dominated by 
small-planetary-mass objects (\cite{sch96}). From a
new theory of hydrodynamic gravitational
condensation (\cite{gib96}$\-- 2000$) these objects
are identified as PFP ``primordial fog particles'' 
that fragmented  within PGC
fragments within proto-galaxy blobs of neutral gas
emerging from the plasma epoch at 300\,000 years. 
The proto-galaxies emerged embedded in a
nested foam of proto-supercluster to proto-galaxy
structures formed previously within the
superviscous Big Bang plasma starting at about
30\,000 years, with the non-baryonic component
diffusing to fill the voids and damp 
gravitational formation of
$\rho$ fluctuations with amplitudes larger than
the $\delta \rho / \rho \approx 8 \times 10^{-5}$
peak level observed,  (\S
\ref{sec1}).  Observations of quasars,
galaxies, and galaxy clusters at large redshifts
support predictions of the theory that
proto-superclusters formed at $z
\approx 5700$, proto-galaxies at $z \approx 1000$,
and the first stars, super-stars, proto-quasars, as
well as the \cite{sch96} ``rogue planets'', near
$z \approx 260$,  (\S
\ref{sec1}). The large population of clumped,
frozen, hydrogenous planetoids predicted can
masquerade as dark energy, quintessence, and the
cosmological constant by causing high
$z$ supernova dimming as an unanticipated time
dependent radiation energy sink and overestimates of
the cluster baryon fraction because of similar
unanticipated gas clumping and cooling effects.
Because the  baryonic matter is mostly sequestered
as PGC clumps of PFPs before the appearance of
quasars, the Gunn-Peterson paradox of the missing
intergalactic gas in quasar spectra is resolved. 

Failure of star-microlensing  studies to observe
equivalent microlensing events from the Bulge and
LMC is not in conflict with the quasar-microlensing
observations because fragmentation of primordial gas
as PGCs and the nonlinear clustering of PFP
fragments within PGCs during their accretion to form
stars produces extremely intermittent lognormal
distributions of the object number density
$n_p$, therefore increasing the minimum number of
independent samples required for statistically
significant estimates of the mean density $\langle
n_p \rangle$ by  star-microlensing far  beyond
practical limits.  
Sample calculations   (\S \ref{sec1}, \cite{gib96},
1997ab, 1999ab, 2000abc) give an estimated 
primordial fog particle (PFP, MBD) condensation mass
range 
$M_{PFP} \approx 10^{-7} \--
10^{-6}  M_{\sun}$,  supporting the
\cite{sch96} quasar-microlensing conclusion that
the  lens galaxy mass of Q0957+561A,B is dominated
by ``rogue planets.'' The EROS/MACHO
star-microlensing  exclusion of such
micro-brown-dwarf objects  as the missing Halo mass
(\cite{alc98}) is attributed  to the
unwarranted assumption that the spatial
distribution of such objects is uniform rather than
extremely intermittent.  

\acknowledgments

\clearpage


\begin{thebibliography}{}

\bibitem[Alcock et al. 1993]{alc93}  Alcock, C., et
al. 1993, \nat, 365, 621

\bibitem[Alcock et al. 1997]{alc97}  Alcock, C. ,
R. A. Allsman, D. Alves, T. S. Axelrod, et al. 
1997, \apj, 486, 697

\bibitem[Alcock et al. 1998]{alc98}  Alcock, C., et
al. 1998, \apjl, 499, L9

\bibitem[Ashman and Zepf 1998]{ash98} Ashman, K.
M.; Zepf, S. E. 1998, Globular Cluster Systems,
Cambridge Univ. Press, Cambridge UK

\bibitem[Auborg et al. 1993]{aub93}  Auborg, E., et
al. 1993, \nat, 365, 623

\bibitem[Arp 1980] {arp80} Arp, H. 1980, Ann. N. Y.
Acad. Sci., 36, 94

\bibitem[Baker and Gibson 1987]{bak87} Baker, M. A.
and C. H. Gibson 1987, J. Phys. Oceanogr., 17:10,
1817

\bibitem[Bershadskii and Gibson 1994]{ber94}
Bershadskii, A. and C. H. Gibson 1994, Physica A,
212:3-4, 251

\bibitem[Binney and Tremaine 1987]{bin87} Binney,
J., and  S. Tremaine 1987, Galactic Dynamics, 
Princeton Univ. Press, Princeton NJ

\bibitem[Burbidge 1979] {bur79} Burbidge, G. 1979,
Nature, 282, 455

\bibitem[Carr 1994] {car94} Carr, B. J. 1994,
ARA\&A, 32, 531

\bibitem[Carr and Sakellariadou 1999] {car99} Carr,
B. J., and M. Sakellariadou 1999, \apj, 516, 195

\bibitem[Chandrasekhar 1951]{cha51} Chandrasekhar,
S. 1951, Proc. Roy. Soc. A, 210, 26-29


\bibitem[Corrigan et al. 1991] {cor91} Corrigan,
R.T.; Irwin, M.J.;  Arnaud, J.; Fahlman, G.G.; et
al. 1991, \aj, 102, 34

\bibitem[De Paolis et al. 1999] {dep99} De Paolis,
F.; Ingrosso, G.; Jetzer, Ph. and Roncadelli, M.
1999,
\apj, 510, L103

\bibitem[De Marchi et al. 1997]{mar97} De Marchi,
G; Clampin, M.; Leitherer, C.; Nota, A.; and Tosi,
M 1997; \apj, 469, L27-L30

\bibitem[De Rujula et al. 1992]{der92} De Rujula,
A. et al.
 1992, \aap, 254, 99

\bibitem[Gaudi and Gould 1997]{gau97} Gaudi, B.S.;
Gould, A.
 1997, \apj, 477, 152

\bibitem[Gibson 1968]{gib68} Gibson, C. H. 1968,
Phys. Fluids, 11, 2305

\bibitem[Gibson 1991]{gib91} Gibson, C. H. 1991,
Proc. Roy. Soc. Lond., A434, 149-164,
astro-ph/9904269

\bibitem[Gibson 1996]{gib96} Gibson, C. H. 1996,
Appl. Mech. Rev., 49, 299, astro-ph/9904260

\bibitem[Gibson 1997a]{gib97a} Gibson, C. H. 1997a,
in The Identification of Dark Matter, Ed.: N. J. C.
Spooner, World Scientific, 114, astro-ph/9904283

\bibitem[Gibson 1997b]{gib97b} Gibson, C. H. 1997b,
in Dark Matter in Astro- and Particle Physics,
Eds.: H. V. Klapdor-Kleingrothaus and Y. Ramachers,
World Scientific, 409, astro-ph/9904284

\bibitem[Gibson 1999a]{gib99a} Gibson, C. H. 1999a,
Journal of Marine Systems, vol. 21, nos. 1-4,
147-167, astro-ph/9904237

\bibitem[Gibson 1999b]{gib99b} Gibson, C. H. 1999b, 
Turbulent mixing, diffusion and gravity in the
formation of cosmological structures: the fluid
mechanics of dark matter, FEDSM99 Conference
proceedings, astro-ph/9904230

\bibitem[Gibson 2000a]{gib00a} Gibson, C. H. 2000a,
Fossils of turbulence and non-turbulence in the 
primordial universe: the fluid mechanics of dark
matter, 8th European Turbulence  Conference,
Barcelona, Spain, June 27-30, Conference 
Proceedings, astro-ph/0002381

\bibitem[Gibson 2000b]{gib00b} Gibson, C. H.
2000b,  Primordial viscosity, diffusivity, Reynolds
numbers, sound and turbulence at the onset of
gravitational structure formation, submitted to
ApJ, astro-ph/astro-ph/9911264

\bibitem[Gibson 2000c] {gib00c} Gibson, C. H.
2000c, Turbulent mixing, viscosity, diffusion and gravity 
in the formation of cosmological structures: the fluid 
mechanics of dark matter, accepted for publication
in the Journal of Fluids Engineering,
astro-ph/0003352 

  
\bibitem[Gibson and Schild 1999a]{gs99a} Gibson, C.
H.  and Schild, R. E. 1999a, in preparation,
astro-ph/9904366

\bibitem[Gibson and Schild 1999b]{gs99b} Gibson,
C. H. and Schild, R. E. 1999b,  Clumps of
hydrogenous planetoids as galactic dark matter,
submitted to ApJ, astro-ph/9908335

\bibitem[Griest 1991]{gri91} Griest, K. 1991, \apj,
366, 412

\bibitem[Griest et al. 1991]{grt91} Griest, K.;
Alcock, C.; Axelrod, T.S.;  Bennett, D.P.; et al.
1991, \apj, 372, L79

\bibitem[Ho and Filippenko 1996] {ho96} Ho,
L. C. and Filippenko, A. V. 1996, \apj, 472, 600

\bibitem[Holtzman et al. 1996]{hol96} Holtzman, J.
A.; Watson, A. M.; Mould, J. R.; Gallagher, J. S.;
Ballester, G. E.; Burrows, C. J.; et al. 1996,
\aj , 112, 416

\bibitem[Irwin et al. 1989]{irw89} Irwin, M.J.;
Webster, R.L.;  Hewett, P.C.; Corrigan, R.T.; et
al., \aj, 98, 1989

\bibitem[Jeans 1902]{jns02} Jeans, J. H. 1902, 
Phil. Trans. R. Soc. Lond. A, 199, 1


\bibitem[Keel 1982]{kee82} Keel, W.C. 1982, \apj,
259, L1

\bibitem[Kerins and Evans 1998]{ker98} Kerins, E.
and Evans, N. 1998,
\apj,  503, L75 

\bibitem[Kundic et al. 1997]{kun97} Kundic, T. et
al. 1997, \apj, 482, 75


\bibitem[Landau and Lifshitz 1959]{lan59} Landau,
L. D. and E. M. Lifshitz 1959, Fluid Mechanics,
Pergamon Press, London

\bibitem[Mushotzky et al. 2000]{mus00} Mushotzky,
R. F.; Cowie, L. L.; Barger, A. J. and Arnaud, K.
A. 2000, Nature, 404, 459

\bibitem[O'Connell et al. 1994]{oco94} O'Connell,
R. W.; Gallagher, J. S. 1994; Hunter, D. A., \apj,
433, 65-79

\bibitem[Paczynski 1986]{pac86} Paczynski, B. 1986, 
\apj, 304, 1


\bibitem[Pelt et al. 1998]{pel98} Pelt, J.; Schild,
R.; Refsdal, S.; Stabell, R. 1998,  A\&A, 336, 829

\bibitem[Refsdal and Stabell 1993]{ref93} Refsdal,
S. and Stabell, R. 1993,
\aap, 278, L5


\bibitem[Renault et al. 1998]{ren98} Renault, C. et
al. 1998,  A\&A, 329, 522

\bibitem[Sahu 1994]{sah94} Sahu, K.C. 1994,
\pasp, 106, 942

\bibitem[Schild 1996]{sch96} Schild, R. 1996, \apj,
464, 125

\bibitem[Schild and Cholfin 1986]{sch86} Schild, R.
and Cholfin, B. 1986, \apj, 300, 209

\bibitem[Schild and Smith 1991]{sch91} Schild, R.
and Smith, R. C. 1991,
 \aj, 101, 813

\bibitem[Schild and Thomson 1994]{sch94} Schild, R.
and Thomson, D.J. 1994, in Gravitational Lenses in
the Universe: proc. 31st Liege Int. Astrophysics
Colloquium, ed. J. Surdej et al. (Liege: Univ.
Liege), 415 

\bibitem[Schild and Thomson 1995]{sch95} Schild, R.
and Thomson, D.J. 1995, AIP CP 336, 95

\bibitem[Schneider et al. 1992]{schn92} Schneider,
P., J.  Ehlers, E.E.  Falco 1992, Gravitational
lenses, Springer-Verlag, New York

\bibitem[Steidel et al. 1999]{sti99} Steidel, C.
C.; Adelberger, K. L.; Giavalisco, M; Dickinson,
M.; and Pettini, M. 1999, \apj, 519, 1

\bibitem[Steidel et al. 2000]{sti00} Steidel, C.;
Adelberger, K.; Shapley, A.; Pettini, M.;
Dickinson, M., and Giavalisco, M. 2000, \apj, 552,
170

\bibitem[Thomson and Schild 1996]{tho96} Thomson,
D.J. and Schild, R.E. 1996, Astrophysical
Applications of Gravitational Lensing,
International Astronomical Union Symposium, 173, 267

\bibitem[Udalski et al. 1993]{uda93}  Udalski, A.,
et al. 1993, Acta Astron., 43, 289

\bibitem[Vanderriest et al. 1989]{van89}
Vanderriest, C.;  Schneider, J.; Herpe, G.;
Chevreton, M.; et al. 1989, \aap , 215, 1

\bibitem[Walker and Wardle 1998]{wal98} Walker, M.,
and Wardle, M. 1998, \apj, 498, L125

\bibitem[Wambsganss et al. 1990]{wam90} 
Wambsganss, J.; Paczynski, B.; Schneider, P. 1990,
\apj , 358, L33

\bibitem[Watson et al. 1996]{wat96} Watson, A. W.;
Gallagher, J. S.; Holtzman, J. A.; Hester, J.;
Mould, J. R.; Ballester, G. E.; et al. 1996, \aj,
112, 534

\bibitem[Weinberg 1972]{win72}  Weinberg, S. 1972,
Gravitation and Cosmology:
 Principles and Applications of the General Theory
of Relativity,  John Wiley and Sons, New York

\bibitem[Whitmore and Schweizer 1995]{whi95}
Whitmore, B. C.; Schweizer, F. 1995, \aj , 109, 960

\bibitem[Yee and De Robertis 1992]{yee92} Yee,
H.K.C.; De Robertis, M.M.   1992, \apj , 398, L21

\bibitem[Zepf et al. 1999]{zep99}Zepf, S. E.;
Ashman, K. M.; English, J.; \& Sharples, R. M.
1999, \aj, 118, 752





\end{thebibliography}
\end{document}